\documentclass[leqno,a4paper,11pt]{article}

\usepackage[colorlinks,
linkcolor=blue,
urlcolor=magenta,
citecolor=red]{hyperref}
\usepackage{amsthm}
\usepackage{amsmath}
\usepackage{amssymb}
 \usepackage{soul}

\usepackage[usenames,dvipsnames]{xcolor}

\usepackage{tikz}
\usetikzlibrary{3d,perspective,calc}
\usepackage{pgfplots}
\pgfplotsset{compat=1.16,width=7cm,height=11cm}

\usepackage{mathtools}
\mathtoolsset{showonlyrefs}

\newcommand{\Con}{\ensuremath{\mathcal{C}}}

\newcommand{\mb}[1]{\ensuremath{\mathbb{#1}}}
\newcommand{\N}{\mb{N}}

\newcommand{\R}{\mb{R}}

\newfont{\bl}{msbm10 scaled \magstep2}

\newcommand{\beq}{\begin{equation}}
\newcommand{\eeq}{\end{equation}}

\newcommand{\notmid}{\mid\kern-0.5em\not\kern0.5em}

\newtheorem{thm}{Theorem}[section]

\newtheorem{defi}[thm]{Definition}

\newcommand{\proj}{\mathrm{proj}}

\newcommand{\LLSn}{Lorentzian length space}
\newcommand{\LpLS}{Lorentzian pre-length space }
\newcommand{\LpLSn}{Lorentzian pre-length space}
\newcommand{\Xll}{\ensuremath{(X,d,\ll,\leq,\tau)} }
\newcommand{\Xlln}{\ensuremath{(X,d,\ll,\leq,\tau)}}

\newcommand{\id}{\mathrm{id}}

\newcommand{\Ric}{\mathrm{Ric}}
\newcommand{\X}{\mathfrak{X}}

\renewcommand{\proj}{\mathrm{pr}}
\renewcommand{\Xll}{$(X,\ll,\leq,\tau)$ }
\renewcommand{\Xlln}{$(X,\ll,\leq,\tau)$}

\renewcommand{\labelenumi}{(\roman{enumi})}
\renewcommand\theenumi\labelenumi

\usepackage{tcolorbox}   
\definecolor{mycolor}{rgb}{0.122, 0.435, 0.698}

\title{A brief introduction to\\ non-regular spacetime geometry}

\author{Clemens S\"amann\thanks{{\tt clemens.saemann@maths.ox.ac.uk} Mathematical Institute, University of Oxford, UK.}}

\begin{document}
 \maketitle

 \begin{abstract}
We give a brief non-technical introduction to \emph{non-regular spacetime geometry}. In particular, we discuss how curvature, and hence gravity, can be defined without a smooth (differential geometric) calculus.
 \end{abstract}

 \tableofcontents

\section{Introduction}
The goal of this short note is to bring two messages across: First, that there are other notions of curvature than the usual differential geometric ones, which generalize to spaces without a differential structure. The second message is the news that such a development is now also possible in the Lorentzian setting (additionally to the Riemannian setting).  Moreover, we discuss in which way these more general notions of curvature are essential for problems in Lorentzian geometry and mathematical General Relativity. This should give a glimpse into a new and very active research area, and might encourage readers to delve deeper (by following the references).

\section{Curvature}\label{sec-cur}
What is curvature? We give a concise technical answer using differential geometry below. However, this is not essential in what is to follow but serves to illustrate that
\begin{enumerate}
 \item to classically define curvature the metric needs to be differentiated twice, and
 \item Einstein's field equations are formulated in this way.
\end{enumerate}

\subsection{Semi-Riemannian geometry}
Here we give a concise review of the differential geometric formulation of curvature for (smooth) semi-Riemannian manifolds. However, this will be only for comparison purposes and can be skipped without problems if one is not familiar with differential geometry. For an introduction to semi-Riemannian geometry see \cite{ONe:83}, which has the advantage that it is geared towards the main applications in General Relativity.
\medskip

On a smooth manifold $M$ we denote by $\X(M)$ the set of all (smooth) vector fields on $M$, i.e., all smooth maps $X\colon M\rightarrow TM$ such that $\proj\circ X = \id_M$, where $\proj\colon TM\rightarrow M$ is the canonical projection and $TM$ is the tangent bundle of $M$.

Let $(M,g)$ be a semi-Riemannian manifold, i.e., a smooth manifold $M$ with a non-degenerate $(0,2)$-tensor $g$ of fixed signature. Usually $g$ is assumed to be smooth but this requirement can be relaxed, for example to $\Con^2$-regularity, which will be enough in what is to follow. In this note we restrict to \emph{Riemannian}, i.e., of signature $(+++\ldots+)$, and \emph{Lorentzian}, i.e., of signature $(-++\dots+)$, metrics. Given a semi-Riemannian metric $g$ we can assign a length to curves as follows:
\begin{equation}
 L_g(\gamma):= \int \sqrt{|g(\gamma',\gamma')|}\,,
\end{equation}
where $\gamma$ is a curve into $M$ of sufficient regularity but can be assumed to be smooth here.

The simplest semi-Riemannian manifolds are the $n$-dimensional \emph{semi-Euclidean spaces} $\R^n_\nu$ of index $\nu\in \{0,1\ldots,n\}$ with metric $\eta$ on $\R^n$ given by
\begin{equation}
 \eta(v,w):= - \sum_{i=0}^\nu v_i w_i + \sum_{i=\nu+1}^n v_i w_i\,,
\end{equation}
where $v=(v_1,\ldots,v_n), w=(w_1,\ldots w_n)\in\R^n$. If the index is zero, i.e., $\nu=0$, then $\R^n_0$ is just $\R^n$ with the usual Euclidean metric and if $\nu=1$, then $\R^n_1$ is $n$-dimensional \emph{Minkowski spacetime} --- the simplest Lorentzian manifold.
\medskip

Given a semi-Riemannian manifold there is a unique \emph{Levi-Civita connection} $\nabla\colon \X(M)\times\X(M)\rightarrow \X(M)$, which has the following properties. 
\begin{enumerate}
 \item The map $X\mapsto \nabla_X Y$ is  $\Con^\infty(M)$-linear for every $Y\in\X(M)$.
 \item The map $Y\mapsto \nabla_X Y$ is $\R$-linear for every $X\in\X(M)$.
 \item It satisfies the \emph{Leibniz rule}, i.e., $\nabla_X(f\, Y) = f\,\nabla_X Y + (X\cdot f)\,Y$ for all $X,Y\in\X(M)$, $f\in\Con^\infty(M$). Here $X\cdot f\in\Con^\infty(M)$ is $p\mapsto T_p f(X(v))$.
 \item It is \emph{torsion-free}, i.e., $\nabla_X \nabla_Y - \nabla_Y \nabla_X  = [X,Y]$ for every $X,Y\in\X(M)$, where $[X,Y](f) = X\cdot (Y\cdot f) - Y\cdot (X\cdot f)$ for $f\in\Con^\infty(M)$ is the commutator of $X$ and $Y$.
  \item Finally, $\nabla$ is \emph{metric}, i.e., it comes from a semi-Riemannian metric: $X \cdot g(Y,Z) = g(\nabla_X Y, Z) + g(Y, \nabla_X Z)$ for all $X,Y,Z\in\X(M)$.
\end{enumerate}

At this point it should be noted that the Levi-Civita connection involves first derivatives of the metric, i.e., to classically make sense of the Levi-Civita connection, the metric $g$ must be at least once (continuously) differentiable. 

The curvature of $(M,g)$ is encoded in the \emph{Riemannian tensor} $R\colon\X(M)^3 \rightarrow \X(M)$, the $(1,3)$-tensor that is given by
\begin{equation}
 R_{X,Y}Z:= \nabla_{[X,Y]} Z - [\nabla_X,\nabla_Y]Z\,.
\end{equation}
Note that the Riemann tensor $R$ involves second derivatives of the metric $g$, which becomes apparent if one writes it in coordinates. Moreover, the Riemann tensor encodes all the curvature information of the semi-Riemann manifold $(M,g)$. In particular, $(M,g)$ is \emph{flat}, i.e., locally isometric to some $\R^n$ if and only if $R=0$.

Another way to fully characterize the curvature of $(M,g)$ is via all the \emph{sectional curvatures} of tangent planes to $M$. Let $p\in M$ and $v,w\in T_pM$ span a tangent plane in $T_pM$, i.e., a non-degenerate two-dimensional linear subspace of $T_pM$. Then the \emph{sectional curvature} $K_p(v,w)$ is defined as
\begin{equation}
 K_p(v,w):= \frac{g(R_{v,w}v,w)}{g(v,v)g(w,w)-g(v,w)^2}\,.
\end{equation}
Note that since the tangent plane is non-degenerate the denominator is non-zero and moreover, $K_p$ does not depend on the choice of $v,w$ spanning the tangent plane.

If one averages the sectional curvatures containing a given unit vector $v\in T_pM$ one obtains the \emph{Ricci curvature} $\Ric(v,v)$, defined as
\begin{equation}
 \Ric_p(v,v):= g(v,v)\,\sum_{i=2}^n K_p(v,e_i)\,,
\end{equation}
where $e_2,\ldots,e_n\in T_pM$ are such that $e_1:=v,e_2,\ldots,e_n$ is an orthonormal basis of $T_pM$ with respect to $g_p$, i.e., $g_p(e_i,e_j) = \epsilon_i \delta_{i,j}$. Here $\delta_{i,j}=1$ if $i=j$, $\delta_{i,j}=0$ otherwise, and $\epsilon_i = g_p(e_i,e_i)=\pm 1$. By polarization and scaling $\Ric$ extends to a $(0,2)$-tensor field on $M$.

There is yet another average (trace) of the Riemann tensor (via the Ricci tensor) that gives the \emph{scalar curvature}, i.e., a scalar function on $M$ that gives an average of the curvatures at $p\in M$, defined as
\begin{equation}
 S_p := \sum_{i\neq j} K_p(e_i,e_j)\,, 
\end{equation}
where $e_1,\ldots,e_n$ is an orthonormal basis of $T_pM$.

\subsection{Gravity is curvature}
Einstein's great insight was that gravity is not a force but the curvature of a Lorentzian manifold. Wheeler succinctly summarized this as
\begin{center}
``Space tells matter how to move. Matter tells space how to curve.''
\end{center}
\cite[p.\ 5]{MTW:73}, \cite[p.235]{Whe:00}. This insight is mathematically expressed by using the Ricci and scalar curvature in the \emph{Einstein field equations} 
\begin{equation}\label{eq-EFE}
 \Ric - \frac{1}{2}S\,g = 8\pi\, T\,,
\end{equation}
where $T$ is the \emph{stress-energy tensor} describing the energy and matter content of the spacetime. So on the left-hand-side of the equation we have the curvature of spacetime, which governs how objects move, and on the right-hand-side the matter, which governs the geometry of spacetime via its curvature. Note that the expression on the left-hand-side, i.e., $\Ric - \frac{1}{2}S\,g$ is the $(0,2)$-tensor containing the same information as the Ricci tensor $\Ric$. In particular, if $T=0$, i.e., if we consider vacuum spacetimes, then the Einstein field equations \eqref{eq-EFE} are equivalent to the \emph{vacuum Einstein equations}
\begin{equation}\label{eq-VEE}
 \Ric = 0\,.
\end{equation}
In conclusion, we see that there is a clear notion of curvature in the differential geometric sense, which relies on sufficient smoothness to define the Riemann tensor, and hence also sectional curvature, Ricci curvature and scalar curvature.

However, from an intuitive point-of-view a space can be curved even if the semi-Riemannian metric is not sufficiently smooth. In fact, one can even imagine spaces that do not come with a metric. So is there another way to detect if one is in a curved space? Yes, there are actually numerous different ways and we will focus on two specific ones in the next section.

\section{Other ways to detect curvature}\label{sec-oth-cur}
Let us start with alternative ways to characterize sectional curvature for Riemannian manifolds $(M,g)$. Recall that a Riemannian metric has signature $(+++\ldots+)$, hence it is positive definite. In particular, the Riemannian metric $g$ gives rise to a canonical metric space structure on $M$. For $x,y\in M$ define
\begin{equation}\label{eq-def-dist}
 d_g(x,y):=\inf\{L_g(\gamma): \gamma \text{ curve that connects } x,y\}\,.
\end{equation}
The metric $d_g$ induces the manifold topology of $M$ and actually captures all the geometry of $(M,g)$ (by a result of Calabi-Hartman \cite{CH:70}, see also \cite{Tay:06}). Another instance of this fact is Toponogov's theorem.

\subsection{Toponogov's theorem}
Toponogov's theorem is an example of so-called \emph{comparison results} in semi-Riemannian geometry. Here, one obtains information on the manifold in question by comparing geometric objects or quantities to simple semi-Rie\-mannian manifolds. The simplest of them are the \emph{spaces of constant curvature}. To be precise, it is enough to consider two-dimensional spaces of constant sectional curvature $K\in\R$. These are the plane $\R^2$ ($K=0$), the scaled sphere $S^2_r$ of radius $r= \frac{1}{\sqrt{K}}$ for $K>0$ and scaled hyperbolic space $\mathbb{H}^2_r$ of radius $r=\frac{1}{\sqrt{-K}}$ for $K<0$. We combine these three cases and write $M_K$ for the two-dimensional Riemannian model space with constant curvature $K\in\R$.

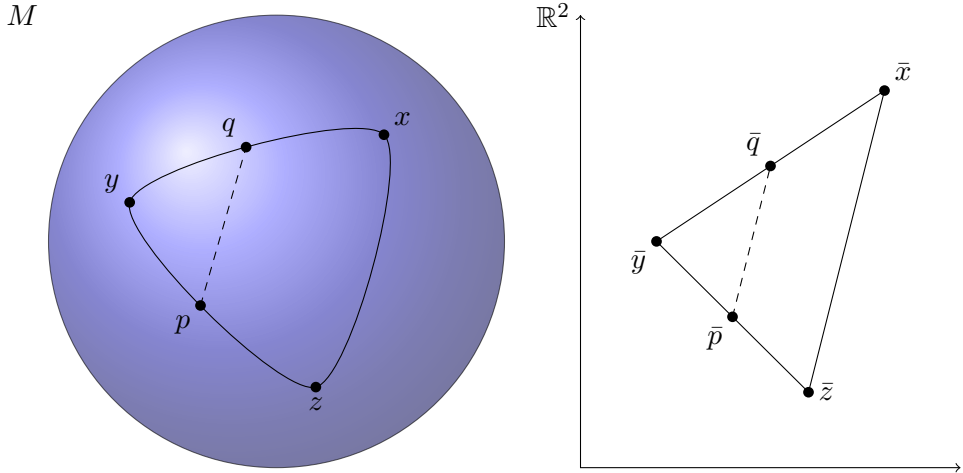
\begin{figure}
 \begin{tikzpicture}[line cap=round,line join=round]
  \shade[draw,ball color=blue!70, opacity=0.6] (0,0,0) circle (3cm);
   \draw node at (-3,3) [left]  {$M$};
   
\coordinate (A) at ({2*cos(45)},{2*sin(45)});
\coordinate (B) at ({2*cos(165)},{2*sin(165)});
\coordinate (C) at ({2*cos(285)},{2*sin(285)});

\fill (A) circle (2pt);
\fill (B) circle (2pt);
\fill (C) circle (2pt);

\draw plot [smooth cycle] coordinates {(A) (B) (C)};

\node at (A) [above right] {$x$};
\node at (B) [above left] {$y$};
\node at (C) [below] {$z$};

\coordinate (P) at (-1,-.85);
\coordinate (Q) at (-.4,1.25);

\fill (P) circle (2pt);
\fill (Q) circle (2pt);

\draw[dashed] (P) -- (Q);

\node at (P) [below left] {$p$};
\node at (Q) [above left] {$q$};

\draw[->] (4,-3) -- (9,-3);
\draw[->] (4,-3) -- (4,3) node[left] {$\R^2$};

\coordinate (a) at (7,-2);
\coordinate (b) at (8,2);
\coordinate (c) at (5,0);

\fill (a) circle (2pt);
\fill (b) circle (2pt);
\fill (c) circle (2pt);

\draw (a) -- (b) -- (c) -- cycle;

\node at (a) [right] {$\bar z$};
\node at (b) [above right] {$\bar x$};
\node at (c) [below left] {$\bar y$};

\coordinate (p) at (6,-1);
\coordinate (q) at (6.5, 1);

\fill (p) circle (2pt);
\fill (q) circle (2pt);

\draw[dashed] (p) -- (q);

\node at (p) [below left] {$\bar p$};
\node at (q) [above left] {$\bar q$};

\end{tikzpicture}
 \caption{Comparing the distance of $p$ to $q$ in the triangle $\Delta xyz$ in $M$ to the corresponding distance of $\bar p$ and $\bar q$ in the comparison triangle $\bar\Delta \bar x\bar y\bar z$ in the plane.}\label{fig-rie-tri-com}
\end{figure}

Let $x,y,z\in M$ be three points in $M$, then together with their distances $d_g(x,y),d_g(y,z), d_g(x,z)$ they form a \emph{triangle} $\triangle xyz$ in $M$. If these distances are small enough compared to $K$ (their sum is smaller than $\frac{2\pi}{\sqrt{K}}$ if $K>~0$), then there is a \emph{comparison triangle} $\triangle\bar x\bar y\bar z$ in the model space $M_K$, see Figure \ref{fig-rie-tri-com}. The comparison triangle has the same side lengths as $\triangle xyz$, i.e., $d_g(x,y) = d_{g_K}(\bar x,\bar y)$, $d_g(y,z) = d_{g_K}(\bar y,\bar z), d_g(x,z) = d_{g_K}(\bar x,\bar z)$, where $d_{g_K}$ is the distance coming from the metric $g_K$ of $M_K$ via \eqref{eq-def-dist}. Such a comparison triangle is unique up to isometries of $(M_K,g_K)$.

With these model geometries at hand it is possible to characterize bounds on the sectional curvature by comparing distances in triangles. We summarize this below in \emph{Toponogov's theorem} (even though Toponogov proved only one direction of the Theorem). For a modern account see e.g.\ \cite[§IX.5]{Cha:06}.

\begin{thm}[Toponogov's theorem \cite{Top:57, Top:58, Top:59}]
A smooth complete Riemannian manifold $(M,g)$ has sectional curvature bounded below by $K\in\R$, i.e., for all $p\in M$: $K_p\geq K$, if and only if for all $\triangle xyz$ with side lengths small enough, $p,q$ on the sides of $\triangle xyz$ and any comparison triangle $\triangle \bar x\bar y\bar z$ in $M_K$ and corresponding points $\bar p,\bar q$ on $\triangle xyz$ we have
\begin{equation}
d_g(p,q)\geq d_{g_K}(\bar{p},\bar{q})\,.
\end{equation}
Here a point $\bar p$ on a side of $\triangle\bar x\bar y\bar z$ \emph{corresponds} to $p$ on a side of $\triangle xyz$ if it has the same distance to the adjacent vertices.
\end{thm}
An illustration for this triangle comparison is Figure \ref{fig-rie-tri-com}. A similar statement holds for sectional curvature bounded above.

Note that Toponogov's theorem only involves the distances of points in $M$ and $M_K$. Thus, one can turn the theorem on its head and \emph{define} bounds on sectional curvature for metric spaces --- without any smooth structure! This is the starting point for the theory of \emph{Alexandrov- and CAT$(K)$-spaces}.

\subsection{Alexandrov- and CAT(K)-spaces}\label{subsec-ale}
First we need to discuss lengths of curves in metric spaces. Let $\gamma\colon[a,b]\rightarrow X$ be a curve (not necessarily continuous) into a metric space $(X,d)$. Then the length of $\gamma$ with respect to $d$ is defined as
\begin{equation}\label{eq-def-len}
 L_d(\gamma):= \sup\{\sum_{i=1}^N d(\gamma(t_{i-1}),\gamma(t_i)): a=t_0 < t_1 < \ldots t_N=b, N\in\N\}\,.
\end{equation}
Now one can define a new metric $\hat d$ analogous to \eqref{eq-def-dist}
\begin{equation}
 \hat d(x,y) := \inf\{L_d(\gamma): \gamma \text{ continuous curve from } x \text{ to } y\}\,,
\end{equation}
and we call $d$ \emph{intrinsic} or $(X,d)$ a \emph{length space} if $\hat d = d$. In a Riemannian manifold $(M,g)$ one has that $L_g = L_{d_g}$, hence the metric space $(M,d_g)$ induced by \eqref{eq-def-dist} is a length space. For an example of a metric space that is not a length space consider the metric induced from $\R^2$ restricted to the circle $S^1$.
\bigskip

At this point we can define \emph{synthetic} sectional curvature bounds for length spaces by turning Toponogov's theorem into a definition. A length space $(X,d)$ has \emph{curvature bounded below} by $K\in\R$ if every point has a neighborhood $U$ such that for all $\triangle xyz$ in $U$ with side lengths small enough, $p,q$ on the sides of $\triangle xyz$ and any comparison triangle $\triangle \bar x\bar y\bar z$ in $M_K$ and corresponding points $\bar p,\bar q$ on $\triangle xyz$ we have
\begin{equation}
d(p,q)\geq d_{g_K}(\bar{p},\bar{q})\,.
\end{equation}
Similarly one defines \emph{curvature bounded above by $K\in\R$}. Neglecting some technicalities the (length) metric spaces with curvature bounded below by some $K\in\R$ are called \emph{Alexandrov spaces} and the ones with curvature bounded above by some $K\in\R$ are called \emph{CAT$(K)$}-spaces (for Cartan-Alexandrov-Toponogov). This is the starting point for \emph{metric geometry}. For further reading see e.g.\ \cite{BBI:01, AKP:19, AKP:23}.

\subsection{Synthetic Ricci curvature bounds}\label{subsec-cd}
\begin{figure}
\begin{center}
 \begin{tikzpicture}[line cap=round,line join=round]
  \shade[draw,ball color=blue!70, opacity=0.5] (5,0) circle (3cm);
   \draw node at (2.6,3) [left]  {$M$};
   
\coordinate (p1) at (5,-2.7);
\coordinate (p2) at (4.9,-2.6);
\coordinate (p3) at (5.2,-2.6);
\coordinate (p4) at (5.1,-2.5);
\coordinate (p5) at (4.96,-2.5);

\fill (p1) circle (1pt);
\fill (p2) circle (1pt);
\fill (p3) circle (1pt);
\fill (p4) circle (1pt);
\fill (p5) circle (1pt);

\coordinate (p11) at (4.8,-1);
\coordinate (p21) at (4.7, -.9);
\coordinate (p31) at (5.1,-.9);
\coordinate (p41) at (5.2,-.8);
\coordinate (p51) at (5,-.8);

\fill (p11) circle (1pt);
\fill (p21) circle (1pt);
\fill (p31) circle (1pt);
\fill (p41) circle (1pt);
\fill (p51) circle (1pt);

\coordinate (p12) at (5,1.7);
\coordinate (p22) at (4.9,1.6);
\coordinate (p32) at (5.2,1.6);
\coordinate (p42) at (5.1,1.5);
\coordinate (p52) at (4.96,1.5);

\fill (p12) circle (1pt);
\fill (p22) circle (1pt);
\fill (p32) circle (1pt);
\fill (p42) circle (1pt);
\fill (p52) circle (1pt);

\draw[dashed] plot [smooth cycle] coordinates {(p1) (p11) (p12)};
\draw[dashed] plot [smooth cycle] coordinates {(p2) (p21) (p22)};
\draw[dashed] plot [smooth cycle] coordinates {(p3) (p31) (p32)};
\draw[dashed] plot [smooth cycle] coordinates {(p4) (p41) (p42)};
\draw[dashed] plot [smooth cycle] coordinates {(p5) (p51) (p52)};
\end{tikzpicture}
 \caption{Positive curvature leads to a convex behaviour of the volume of clouds of points --- first they spread out and then they refocus.}\label{fig-OT}
 \end{center}
\end{figure}
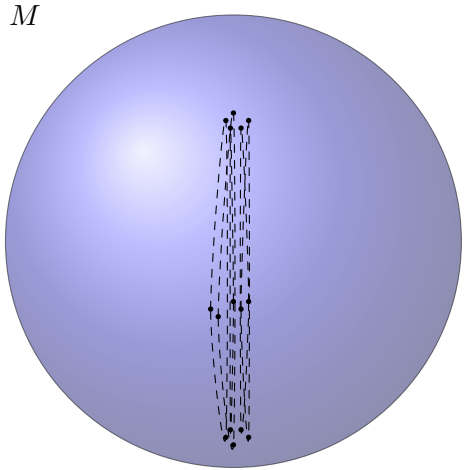

There is also a synthetic notion of Ricci curvature bounds in the Riemannian setting pioneered independently by Lott-Villani and Sturm \cite{LV:09,Stu:06a, Stu:06b}. Introducing this approach in more detail would go beyond the scope of this brief article but we outline the main idea. In metric spaces with curvature bounded below or above (Subsection \ref{subsec-ale}) one compares geodesics (minimizing curves) in the space to ones in the model spaces. As Ricci curvature is a kind of an average of sectional curvatures one cannot expect to control the behaviour of all the geodesics. Instead, we want to contrast the behaviour of ``almost'' all geodesics with a comparison situation in model spaces. Thus, we need another ingredient to quantify what ``almost'' should mean and one does this by introducing a \emph{reference measure}, i.e., a \emph{Radon measure} $\mathfrak{m}$ on $(X,d)$. This is summarized in the notion of a \emph{metric measure space} $(X,d,\mathfrak{m})$ and is the start of \emph{metric measure geometry}. The crucial ingredient to make this whole theory work is \emph{optimal transport}, cf.\ e.g.\ \cite{Vil:09}, which goes back more than two hundred years to the work of Monge. It was then reinvigorated by Kantorovich in the last century and the central idea is to find the \emph{optimal} (i.e., cheapest) way of transporting goods, material or similar from one location to another. Clearly, the cost of transporting something is correlated with the distance of the source and the destination. However, geometry influences the cost as well: it might be cheaper to go around a mountain than up and down. This provides yet another way to detect curvature! Optimal transport works on quite general metric measure spaces --- again there is no need for a manifold or smooth structure, and thus, one can define curvature bounds by comparing how clouds of points in the space are optimally transported to a comparison situation in model spaces. This amounts to convexity/concavity properties of entropies of these clouds of points, i.e., how does the volume of the clouds of points behave when the points are moving in the optimal way, see Figure \ref{fig-OT}. This is summarized in the so-called \emph{curvature-dimension condition}, and a metric measure space $(X,d,\mathfrak{m})$ is a $\mathsf{CD}(K,N)$-space if it has Ricci curvature bounded below by $K\in\R$ (in the optimal transport sense) and dimension bounded above by $N\in(0,\infty]$. In fact, a negative dimensional bound is also possible and useful \cite{Oht:16}.

\bigskip
The introduction of Alexandrov-, CAT$(K)$- and $\mathsf{CD}(K,N)$-spaces has had a huge impact on smooth Riemannian geometry as it allows one to take limits of Riemannian manifolds, which might not be smooth Riemannian manifolds anymore. In particular, one obtains \emph{precompactness} theorems \cite[Prop.\ 5.2]{Gro:99} and this in turn allows one to say something about smooth Riemannian manifolds, e.g.\ \cite{CC:97}. So, even if one is only interested in smooth classical Riemannian geometry one is led to study such limits which are metric spaces with additional structure (e.g.\ length spaces with synthetic sectional curvature bounds, metric measure spaces with synthetic Ricci curvature bounds, etc.). For a recent review of synthetic Ricci curvature bounds see \cite{Stu:24}.

\section{Lorentzian geometry}
While the focus of Section \ref{sec-cur} was more on the general semi-Riemannian case, and later only on the Riemannian case, we focus here exclusively on the Lorentzian case, as it differs significantly. Here we briefly recall its main ingredients.

Let $(M,g)$ be a Lorentzian manifold. Then we can classify tangent vectors as follows. Let $p\in M$, then $v\in T_p M$ is called

\begin{equation}
\begin{cases}
 \text{\emph{timelike}}\\
 \text{\emph{null}}\\
 \text{\emph{causal}}\\
 \text{\emph{spacelike}}
 \end{cases}
 \text{\quad if \qquad}
 g_p(v,v)\qquad
 \begin{cases}
  < 0\,,\\
  = 0 \text{ and } v\neq 0\,,\\
  \leq 0\text{ and } v\neq 0\,,\\
  >0\text{ or } v=0\,.
 \end{cases}
\end{equation}

\begin{figure}
 \begin{center}
  \includegraphics[scale=.2]{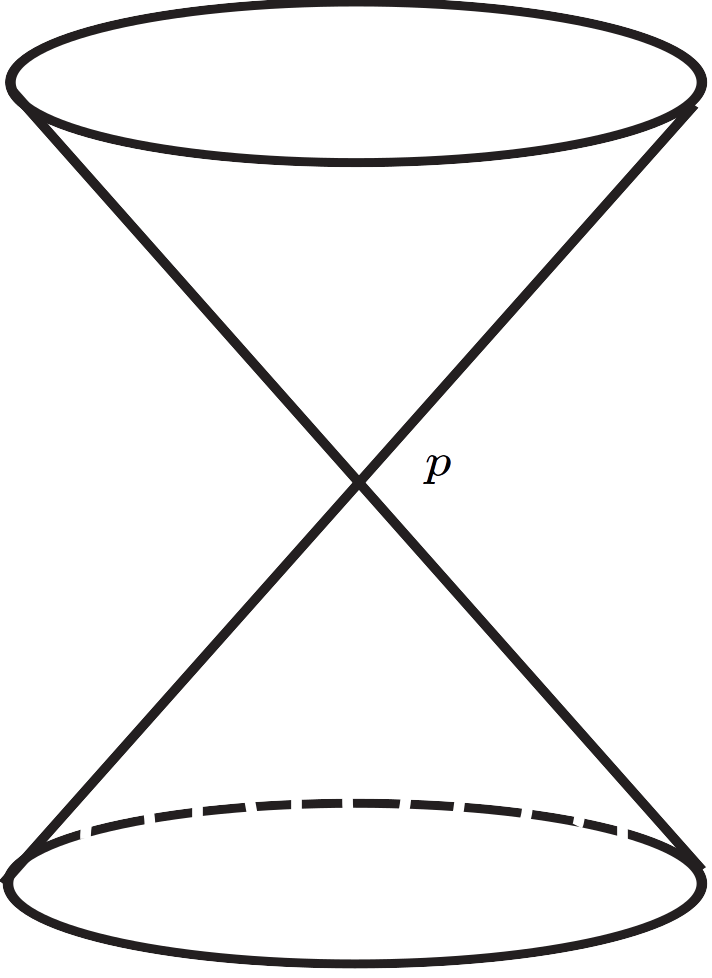}
 \end{center}
\caption{Causal cone at $p$.}\label{fig-lig-con}
\end{figure}

What makes the Lorentzian case unique among all signatures is that the \emph{light cone} in $T_pM$ i.e., $\{v\in T_p M: v$ causal$\}$, consists of two connected components, see Figure \ref{fig-lig-con}.
Thus at every point $p\in M$ we have a choice of what we call the causal \emph{future} of $p$ and the causal \emph{past} of $p$. When one can make such a choice continuously and globally we say that the Lorentzian manifold is \emph{time orientable}. A time orientation is for example given by a global timelike vector field $T\in\X(M)$, i.e., $g_p(T(p),T(p))<0$ for all $p\in M$. Then, we say that a causal vector $v\in T_pM$ is \emph{future directed} if $g_p(v,T(p))<0$ and \emph{past directed} if $g_p(v,T(p))>0$. Analogously, one defines future-/past-directed timelike/null/causal/spacelike curves into $M$ by requiring their tangent vectors to be of this class.

A connected time oriented Lorentzian manifold is called \emph{spacetime} and this is the stage and the central object of \emph{General Relativity}, Einstein's theory of gravity. The points in a spacetime $(M,g)$ are called \emph{events}, as they represent an instance of time and a position in space, but unlike in Newtonian physics (or Special Relativity) there is no global notion of time (coordinate) and spacetime does not necessarily split into space {\bf and} time. The events in the causal future of a point $p$ can be influenced from $p$ by signalling or travelling with at most the speed of light. Thus, as speeds greater than the speed of light are not possible in General Relativity, events outside of the causal future cannot be influenced by $p$. Similarly, events in the timelike future of $p$ can only be influenced by travelling slower than light speed (\emph{subluminal}). We can capture these relations in the \emph{timelike} $\ll$ and \emph{causal relations} $\leq$. We say that $p\ll q$ ($p< q$) if there is a future directed timelike (causal) curve from $p$ to $q$, and $p\leq q$ if $p<q$ or $p=q$.

An \emph{observer} $\gamma$ in a spacetime is a future or past directed causal curve so that no forces act on it, i.e., it is (up to parametrization) a causal \emph{geodesic}. This means that it solves
\begin{equation}
 \nabla_{\gamma'}\gamma'=0\,,
\end{equation}
and consequently, it \emph{maximizes} the time the observer experiences. This can be formulated via the so-called \emph{time separation function} (or sometimes \emph{Lorentzian distance}):
\begin{equation}\label{eq-def-tau}
 \tau(p,q):=\sup \{L_g(\gamma): \gamma \text{ future directed causal from }p \text{ to } q\}\cup\{0\}\,.
\end{equation}
Then a causal geodesic locally maximizes $\tau$ between points on it and if a curve maximizes $\tau$ then it is (up to parametrization) a causal geodesic.

The definition of the time separation function is analogous to the distance in the Riemannian case \eqref{eq-def-dist} but there are significant differences. First, $\tau$ is not a metric. In fact, it satisfies the \emph{reverse triangle inequality}
\begin{equation}
 \tau(p,q) + \tau(q,r) \leq \tau(p,r)\qquad \forall p\leq q\leq r\,.
\end{equation}
Moreover, $\tau$ is not positive definite and not symmetric. For example $\tau(p,q)=0$ for all points $p\not\leq q$. However, $\tau$ still encodes the geometry of the spacetime as the following celebrated result of Hawking, King and McCarthy shows (see also \cite[Thm.\ 4.17]{BEE:96}).

\begin{thm}[Hawking, King, McCarthy \cite{HKM:76}]
 Let $(M,g)$, $(N,h)$ be two spacetimes of the same dimension and suppose that $(M,g)$ has no almost closed causal curves (it is \emph{strongly causal}). If there exists a time separation preserving map $\phi\colon M\rightarrow N$, i.e.
 \begin{equation}
  \tau_h(\phi(p),\phi(q)) = \tau_g(p,q)\qquad \forall p,q\in M\,,
 \end{equation}
then $\phi$ is actually a smooth isometry of $(M,g)$ and $(N,h)$.
\end{thm}
This result should be viewed as the Lorentzian counterpart of the Calabi-Hartman \cite{CH:70} result mentioned in Section \ref{sec-oth-cur}. So if the time separation function $\tau$ encodes the geometry of (strongly causal) spacetimes, can we develop a geometry built on this spacetime distance in the same way as Alexandrov- and CAT$(K)$-spaces are built on the metric distance? In particular, every Riemannian manifold is canonically a metric space but a spacetime is not. So what should the analog of a metric space be in the Lorentzian setting? We provide an answer in the next section.

\section{Non-regular Lorentzian geometry}
Our suggestion for an analog for metric spaces in the Lorentzian setting are \emph{\LpLSn s} \cite{KS:18}:

\begin{defi}[{\cite[Def.\ 2.8]{KS:18}}]
Let $X$ be a metrizable topological space, $\leq$ a preorder on $X$, $\ll$ a transitive relation contained in $\leq$, and $\tau\colon X\times X\rightarrow[0,\infty]$  lower semicontinuous. Then \Xll is a \emph{Lorentzian pre-length space} if
\begin{equation}
 \tau(x,z)\geq \tau(x,y) + \tau(y,z)\qquad \qquad \forall x\leq y\leq z\,,
\end{equation}
and $\tau(x,y)=0$ if $x\nleq y$, as well as $\tau(x,y)>0$ if and only if $x\ll y$. We call $\tau$ the \emph{time separation function}.
\end{defi}

Natural examples of \LpLSn s are spacetimes with their causal relations and time separation function \eqref{eq-def-tau}. Moreover, directed graphs can also be viewed as \LpLSn s, where the time separation function is given by counting the maximum number of steps one needs to take to get from one vertex to another. This provides a first example outside of smooth Lorentzian geometry \cite[Ex.\ 2.16]{KS:18}.

Let $\gamma$ be a curve from an interval $I$ into a \LpLS \Xlln, which need not be continuous. Then $\gamma$ is called
\begin{equation}
\begin{cases}
 \text{\emph{timelike}}\\
 \text{\emph{causal}}\\
  \text{\emph{null}}
 \end{cases}
 \text{\quad if \qquad}
 \begin{cases}
  \gamma(s)\ll \gamma(t)\,,\\
  \gamma(s)\leq \gamma(t)\,,\\
  \gamma(s)\leq \gamma(t)\ \&\ \gamma(s)\not\ll\gamma(t)\,,
 \end{cases}
\end{equation}
for all $s,t\in I$ with $s\leq t$.  Note that such curves are implicitly future directed as they are defined via the timelike and causal relations. This notion of causal character for curves does not necessarily agree with the usual notion in spacetimes, but if the spacetime is strongly causal, then the two notions of causal curves agree \cite[Lem.\ 2.21]{KS:18}. This does not hold for timelike curves cf.\ e.g.\ \cite[Ex.\ 2.20]{KS:18}. However, one can define a \emph{length} $L_\tau$ of causal curves that agrees with the $g$-length of curves in a smooth spacetime $(M,g)$ \cite[Prop.\ 2.32]{KS:18}. This length $L_\tau(\gamma)$ of a causal curve $\gamma\colon[a,b]\rightarrow X$ is defined as
\begin{equation}
 L_\tau(\gamma) := \inf \{\sum_{i=1}^N \tau(\gamma(t_{i-1}),\gamma(t_i)): a=t_0 < t_1 < \ldots t_N=b, N\in\N\}\,.
\end{equation}
Again, note that this is analogous to the metric case \eqref{eq-def-len}. Having a notion of length, we can now define what causal geodesics should be: They should \emph{maximize} the $\tau$-length between their endpoints, i.e., a causal curve $\gamma\colon[a,b]\rightarrow X$ is a \emph{causal geodesic} or a \emph{maximizer} if
\begin{equation}
 \tau(\gamma(a),\gamma(b)) = L_\tau(\gamma)\,.
\end{equation}
Finally, we can now also say when the given time separation $\tau$ is \emph{intrinsic}, i.e., introduce the Lorentzian analog of metric length spaces. Given a \LpLS \Xll we can define a new time separation function as one would define it in spacetimes, i.e., cf.\ \eqref{eq-def-tau}
\begin{equation}
 \hat\tau(x,y):=\sup\{L_\tau(\gamma): \gamma \text{ causal curve from }x \text{ to } y\}\cup\{0\}\,.
\end{equation}
Then $\tau$ is \emph{intrinsic} if $\hat\tau = \tau$ and \Xll is a \emph{Lorentzian length space} if $\tau$ is intrinsic and some additional technical conditions hold \cite[Def.\ 3.22]{KS:18}.
\bigskip

To define curvature bounds for \LpLS we first need to know to which model spaces we want to compare to. These are the two-dimensional (simply connected) Lorentzian spaces of constant curvature: Two-dimensional Minkowski space $\R^2_1$ ($K=0$), the universal cover of scaled de Sitter spacetime ($K>0$), and the universal cover of scaled anti-de Sitter spacetime ($K<0$). To be precise, $L_K$ is defined as
\begin{equation}
L_K:=
\begin{cases}
 \R^2_1\\
 \tilde S^2_1(\frac{1}{\sqrt{K}})\\
 \tilde H^2_1(\frac{1}{\sqrt{-K}})\\
 \end{cases}
 \text{\quad if \qquad}
 \begin{cases}
  K=0\,,\\
  K>0\,,\\
  K<0\,,
 \end{cases}
\end{equation}
where $\tilde S^2_1(r)$ is the simply connected Lorentzian covering manifold of $S^2_1(r):=\{v\in\R^3_1: \eta(v,v) = r^2 \}$ and $\tilde H^2_1(r)$ is the simply connected Lorentzian covering manifold of $H^2_1(r):=\{v\in\R^3_2: \eta(v,v)=-r^2 \}$ (for $r>0$).

We say that a \LpLS \Xll has \emph{timelike curvature bounded below by $K\in\R$} if every point $p\in X$ has a neighborhood $U$ such that
\begin{enumerate}
 \item the time separation function is finite-valued and continuous on $U\times U$,
 \item for all $x,y\in U$ with $x\leq y$ there is a maximizer from $x$ to $y$, and
 \item for all $x\ll y \ll z$, $x,y,z\in U$ and $p,q$ on the sides of the timelike triangle $\Delta xyz$, $\bar p,\bar q$ corresponding points in the comparison triangle $\Delta\bar x\bar y\bar z$ of $\Delta xyz$ we have that
 \begin{equation}
  \tau(p,q)\leq \bar\tau(\bar p,\bar q)\,,
 \end{equation}
where $\bar\tau$ is the time separation function of $L_K$.
\end{enumerate}
Note that for timelike curvature bounded below the time separation of the points in the triangle is actually bounded \emph{above}. Moreover, the first two points only ensure that there is actually something to compare and only the third point is the triangle comparison condition. Analogously, one defines timelike curvature bounded above. Similar to Toponogov's theorem, this triangle comparison condition is equivalent to bounds on the timelike sectional curvatures of smooth (strongly causal) spacetimes:

\begin{thm}[Beran, Kunzinger, Ohanyan, Rott {\cite[Thm.\ 3.1, Thm.\ 3.2]{BKOR:24}}]
 Let $(M,g)$ be a smooth strongly causal spacetime. Then $(M,g)$ has timelike curvature bounded below (or above) by $K\in\R$ if and only if for all $p\in M$, $v,w$ spanning a timelike tangent plane at $p$ we have
 \begin{equation}
    K_p(v,w) \leq K \quad \text{ (or } K_p(v,w)\geq K \text{)}\,.
 \end{equation}
\end{thm}
So the inequalities on the sectional curvatures of the tangent planes flip as compared to the Riemannian case. Moreover, such an equivalence of one-sided sectional curvature bounds hold for general (smooth) semi-Riemannian manifolds if one also includes spacelike sectional curvatures appropriately \cite{AB:08}.
\bigskip

This notion of curvature bounds opened up the possibility to talk about curvature for spacetimes of low regularity, for example even for spacetimes where the Lorentzian metric is only continuous and might not be differentiable anywhere \cite[Subsec.\ 5.1]{KS:18}, \cite{Lin:23}. In fact, no differentiable or manifold structure is needed at all, which comes up naturally in gluing constructions \cite{BR:24, Rot:23} and \emph{causal boundaries} \cite{ABS:22}. This is a significant advantage to classical Lorentzian geometry and General Relativity. Another advantage is that one can make sense of curvature blow-up in situations where this is classically impossible, for example in non-smooth extensions of spacetimes. This was leveraged in \cite{GKS:19} to extend \cite{GLS:18} and, in particular, to relate inextendibility with a blow-up of curvature --- which had not been possible before. This is an instance of the \emph{paradigm} that using the synthetic framework of \LLSn s one can obtain far reaching conclusions about smooth, classical spacetimes. Moreover, we provide a large class of examples of \LLSn s in \cite{AGKS:21} and give a first synthetic singularity theorem. Further, there are by now several formulations of the synthetic timelike curvature bounds \cite{BKR:23}, and some of them even make sense for discrete spaces. Additionally, globalization of these local curvature bounds was established in \cite{BNR:23,BHNR:23} and a splitting theorem for globally hyperbolic \LLSn s with non-negative timelike curvature
was given in \cite{BORS:23}.
\medskip

So far, we only discussed synthetic analogs of timelike sectional curvature bounds, but as we have seen in Section \ref{sec-cur} Ricci curvature is essential in General Relativity. Independently, McCann and Mondino-Suhr characterized timelike Ricci curvature bounds for smooth spacetimes \cite{McC:20, MS:23} using techniques from optimal transport (as outlined in Subsection \ref{subsec-cd} in the Riemannian case). Building on these works and using our framework of \LpLSn s Cavalletti-Mondino introduced a notion of synthetic timelike Ricci curvature bounds, $\mathsf{TCD}$-spaces as an analog to the metric $\mathsf{CD}$-spaces, and proved a synthetic Hawking singularity theorem \cite{CM:20}. As compared to the Riemannian setting, now the ``cost'' is given by the time separation function and the cloud of points can only be transported with subluminal speed --- hence along future directed timelike maximizers (geodesics). Different formulations of synthetic timelike Ricci curvature bounds are given in \cite{Bra:23b}, synthetic formulations of the \emph{null energy condition} from General Relativity in \cite{Ket:24, McC:24}, and variable timelike Ricci curvature bounds in \cite{BMcC:23}. Finally, Mondino-Suhr also give a synthetic formulation of the vacuum Einstein equations \eqref{eq-VEE} \cite[App.\ B]{MS:23} and Cavalletti-Mondino establish timelike isoperimetric inequalities, which include even new ones in the smooth spacetime case \cite{CM:24}.

\section{Conclusion}
The list of works given in the previous section is not exhaustive but for space restrictions some choices had to be made. Therefore, let me direct you also to the recent review \cite{CM:22}. Furthermore, the above shows that we are here only at the beginning of a new and exciting development and that there are many open questions in this very active field. Finally, let me point out that non-regular spacetime geometry has found surprising applications even outside of spacetime geometry. For example, it found use in other areas of geometry (e.g.\ \cite{Hed:22}) and even in machine learning \cite{LL:23}. This seems to indicate that Lorentzian (pre-)length spaces capture the fundamentals of Lorentzian geometry but at the same time are sufficiently flexible to be applicable to quite different settings.

\subsection*{Acknowledgement}
The author thanks Michael Kunzinger, Argam Ohanyan and Roland Steinbauer for valuable comments on the draft.

\noindent C.S.\ is supported by the European Research Council (ERC), under the European’s Union Horizon 2020 research and innovation programme, via the ERC Starting Grant “CURVATURE”, grant agreement No.\ 802689.

\bibliographystyle{halpha-abbrv}
\bibliography{Master}
\addcontentsline{toc}{section}{References}

\end{document}